\documentclass[iop]{emulateapj}
\usepackage[version=3]{mhchem}
\usepackage{longtable}
\usepackage{apjfonts}

\makeatletter
\newcounter{reaction}
\renewcommand\thereaction{C\,\arabic{reaction}}
\newcommand\reactiontag{\refstepcounter{reaction}\tag{\thereaction}}
\newcommand\reaction@[2][]{\begin{equation}\ce{#2}%
\ifx\@empty#1\@empty\else\label{#1}\fi%
\reactiontag\end{equation}}
\newcommand\reaction@nonumber[1]{\begin{equation*}\ce{#1}%
\end{equation*}}
\newcommand\reaction{\@ifstar{\reaction@nonumber}{\reaction@}}
\makeatother

\shorttitle{Visible Phase Curves of Exoplanets}

\shortauthors{Hu et al.}

\begin{document}

\title{A Semi-Analytical Model of Visible-Wavelength Phase Curves of Exoplanets and Applications to Kepler-7~b and Kepler-10~b}

\author{Renyu Hu$^{1,2,*}$, Brice-Olivier Demory$^3$, Sara Seager$^{4,5}$, Nikole Lewis$^{4,+}$, and Adam P. Showman$^{6}$}
\affil{$^1$Jet Propulsion Laboratory, California Institute of Technology, Pasadena, CA 91109}
\affil{$^2$Division of Geological and Planetary Sciences, California Institute of Technology, Pasadena, CA 91125}
\affil{$^3$Astrophysics Group, Cavendish Laboratory, J.J. Thomson Avenue, Cambridge CB3 0HE, UK.}
\affil{$^4$Department of Earth, Atmospheric and Planetary Sciences, Massachusetts Institute of Technology, Cambridge, MA 02139}
\affil{$^5$Department of Physics, Massachusetts Institute of Technology, Cambridge, MA 02139}
\affil{$^6$Department of Planetary Sciences, University of Arizona, Tucson, AZ 85721}
\email{$^*$Hubble Fellow \\ $^+$Sagan Fellow \\ Email: renyu.hu@jpl.nasa.gov \\ Copyright 2015. All rights reserved.}

\begin{abstract}
{\it Kepler} has detected numerous exoplanet transits by precise measurements of stellar light in a single visible-wavelength band. In addition to detection, the precise photometry provides phase curves of exoplanets, which can be used to study the dynamic processes on these planets. However, the interpretation of these observations can be complicated by the fact that visible-wavelength phase curves can represent both thermal emission and scattering from the planets. Here we present a semi-analytical model framework that can be applied to study {\it Kepler} and future visible-wavelength phase curve observations of exoplanets. The model efficiently computes reflection and thermal emission components for both rocky and gaseous planets, considering both homogeneous and inhomogeneous surfaces or atmospheres. We analyze the phase curves of the gaseous planet Kepler-7~b and the rocky planet Kepler-10~b using the model. In general, we find that a hot exoplanet's visible-wavelength phase curve having a significant phase offset can usually be explained by two classes of solutions: one class requires a thermal hot spot shifted to one side of the substellar point, and the other class requires reflective clouds concentrated on the same side of the substellar point. The two solutions would require very different Bond albedos to fit the same phase curve; atmospheric circulation models or eclipse observations at longer wavelengths can effectively rule out one class of solutions, and thus pinpoint the albedo of the planet, allowing decomposition of the reflection and the thermal emission components in the phase curve. Particularly for Kepler-7~b, reflective clouds located on the west side of the substellar point can best explain its phase curve. We further derive that the reflectivity of the clear part of the atmosphere should be less than 7\% and that of the cloudy part should be greater than 80\%, and that the cloud boundary is located at $11\pm3$ degree to the west of the substellar point. For Kepler-10~b, the phase curve does not show a significant phase offset, and any model with a Bond albedo greater than 0.8 would provide an adequate fit. We suggest single-band photometry surveys could yield valuable information on exoplanet atmospheres and surfaces. 
\end{abstract}


\section{Introduction}

A great number of exoplanets have been discovered by precise photometry.  NASA's {\it Kepler} spacecraft, monitoring 160,000 stars in the sky, discovered that more than half of the stars should host planets smaller than Neptune (e.g., Fressin et al. 2013; Howard 2013). The CHaracterizing ExOPlanet Satellite (CHEOPS), the Transit Exoplanet Sky Survey (TESS), and the PLAnetary Transits and Oscillations of stars (PLATO), designed to search for exoplanets around nearby bright stars using the same technique as {\it Kepler}, have recently been selected by ESA and NASA for launch within the next decade. The precise measurements of light curves of stars in a single visible-wavelength band will continue to be a dominant way to detect exoplanets, especially rocky exoplanets, in the coming years.

Beyond planet detection, {\it Kepler} and future transiting exoplanet search missions may also provide valuable information on the nature of exoplanets via measuring the planets' phase curves. The phase curves reveal longitudinal information regarding the planets' atmosphere or surface. At visible wavelengths, a phase curve could illustrate how a planet reflects stellar light and provide an effective way to study the condensed-phase particles in the planet's atmosphere or the planet's surface (e.g., Madhusudhan \& Burrows 2012; Hu et al. 2013; Heng \& Demory 2013; Demory et al. 2013). The short-wavelength wing of thermal emission may extend to the visible wavelengths and affect the phase curve if the planet is hot enough  (e.g., Rouan et al. 2011). It is therefore useful to learn whether a phase curve of combined reflection and thermal emission can place constraints on the atmosphere or surface of an exoplanet.

The phase curve at visible wavelengths has been a powerful diagnostic tool to characterize Solar-System bodies, a method orthogonal to analyzing spectral features. The center-to-limb variation of the reflectivity continuum and major spectral features of Jupiter has yielded detailed information on the location and layering of its clouds (Sato \& Hanson 1979). The phase curve of Venus alone has provided strong constraints on the refractive index of its cloud particles (Arking \& Potter 1968), which were later improved to effectively only allow spherical sulfate particles by measuring polarization (Hansen \& Hovenier 1974). The photometric light curves of asteroids have been used extensively to study their shape and rotational properties (e.g., Torppa et al. 2003). 

For exoplanets, the phase curve characterization has been mostly limited to mid-infrared wavelengths due to greater planet-to-star flux ratios than at visible wavelengths (e.g., Knutson et al., 2007, 2009a,b, 2012; Cowan et al. 2007, 2012; Crossfield et al. 2010; Lewis et al. 2013; Maxted et al. 2013; Zellem et al. 2014). Many mid-infrared phase curves of exoplanets show phase peaks before secondary eclipses, which have been interpreted as evidence for eastward displaced hot spots driven by super-rotating planet-encircling jet streams in the atmospheres (e.g., Showman et al. 2008, 2009). The visible-wavelength phase curves have been observed for one Jupiter-sized exoplanet before {\it Kepler} (Snellen et al. 2009), and recently for a number of exoplanets with {\it Kepler} data (e.g., Demory et al. 2013; Esteves et al. 2014).

The interpretation of visible-wavelength phase curves without any spectral information, as is the case for {\it Kepler} observations and also expected for future space photometry missions, are likely to be complicated by the following two factors: (1) a visible-wavelength phase curve that contains reflected light will depend on the spatial distribution of clouds on the planet; (2) a visible-wavelength phase curve may contain both reflection and thermal emission from the planet, if the planet is highly irradiated.  

To interpret the visible-wavelength phase curves of exoplanets, we construct a simple semi-analytical model that can be applied to both {\it Kepler} and future visible-wavelength phase curve observations of exoplanets. Models at different levels of sophistication have been developed to study atmospheric circulation on exoplanets and interpret their thermal phase curves  (e.g., Showman \& Guillot 2002; Showman et al. 2008, 2009; Rauscher \& Menou 2010, 2012; Castan \& Menou 2011; Wordsworth et al. 2011; Heng et al. 2011a, b; Cowan \& Agol 2011; Heng \& Kopparla 2012; Perna et al. 2012; Showman \& Kaspi 2013; Showman et al. 2013; Dobbs-Dixon \& Agol 2013; Parmentier et al. 2013; Mayne et al. 2014; Rauscher \& Kempton 2014; Kataria et al. 2014; Showman et al. 2014). However, because the {\it Kepler} observations contain only a single band, it is impractical to constrain the composition of the planet's atmosphere or surface. We approach the problem based on a physically motivated parameterization: instead of studying the detailed physical processes, we try to constrain several overarching physical parameters from the observations. This way we can compare planetary scenarios and shed light on general questions such as, whether the planet has an atmosphere, and whether the atmosphere has a standing circulation pattern and/or patchy clouds.

This paper is organized as follows. We briefly outline possible planetary scenarios important for phase curves in \S~2, and then describe our semi-analytical model for interpreting the visible-wavelength phase curve of exoplanets in \S~3. In \S~4 we apply our interpretation tool to study the phase curve of the gaseous planet Kepler-7~b, and in \S~5 we apply our interpretation tool to study the phase curve of the rocky planet Kepler-10~b. We discuss model degeneracies and suggest how they could be addressed with additional observations in \S~6 and conclude in \S~7.

\section{Planetary Scenarios}

\begin{figure*}
\begin{center}
 \includegraphics[width=1.0\textwidth]{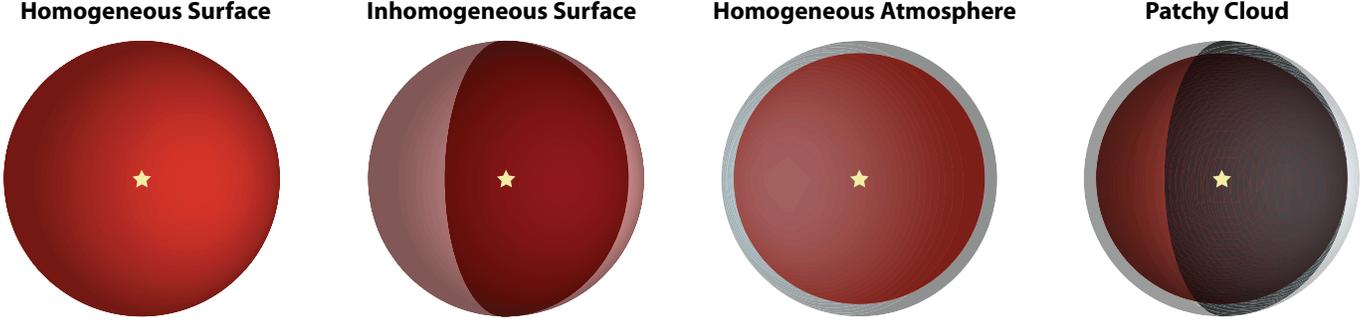}
 \caption{
Schematic illustration of the four planetary scenarios important for visible-wavelength phase curves.
The figure shows the planet's dayside, and the star mark indicates the location of the substellar point.
In the homogeneous surface  scenario, the planet is covered by a solid surface or a molten lava ocean, and does not have an atmosphere.
In the inhomogeneous surface scenario, the planet has a surface containing patches of differing albedo and/or thermal
inertia, and does not have an atmosphere.
One example of the inhomogeneous surface scenarios, as depicted on the figure, is that a fraction of the dayside is covered by molten lava (dark red) and the rest is covered by solid surface (light red). The solid part can be more reflective than the molten part.
In the homogeneous atmosphere scenario, the planet has an atmosphere free of clouds or fully covered by clouds.
In the patchy cloud scenario, a fraction of the dayside is cloud-free (dark grey) and the rest is cloudy (light grey). The cloudy part is more reflective than the cloud-free part.
 }
 \label{scenario}
  \end{center}
\end{figure*}

To make the model applicable for both gaseous and rocky exoplanets, we consider the following four planetary scenarios (see Figure \ref{scenario} for a schematic illustration). These four scenarios are separated by whether a planet has an atmosphere, and whether a planet has a homogeneous reflecting layer. Whether a planet has an atmosphere is important because the atmosphere could cause the greenhouse effect and raise the emitting temperature of the planet. Whether a planet has a homogeneous reflecting layer is also important because an inhomogeneous reflecting layer may cause a phase curve offset.

The first and second scenarios are for bare-rock planets without atmospheres, and they differ by whether the surface is homogeneous. In the ``homogenous surface'' scenario, the planet is covered by a solid surface or a molten lava ocean. In the ``inhomogeneous surface'' scenario, the planet's surface contains patches of differing albedo and/or thermal inertia. One example of the inhomogeneous surface scenarios is that only parts of the planet's dayside are molten, forming a lava lake (e.g., L\'eger et al. 2011). The molten surface and the solid surface may have different reflectivities. A key feature that makes this scenario relevant is that the lava lake does not need to extend symmetrically with respect to the substellar point. This is an analog of a recent simulation on the ice coverage of tidally-locked ocean planets around M dwarf stars (Hu \& Yang 2014). If a molten surface has a lower reflectivity than a solid surface of the same material, an asymmetric lava lake may lead to a phase shift via reflection or thermal emission. 

The third and fourth scenarios are for planets with atmospheres, and they differ by whether the atmosphere has patchy clouds. The distribution of the condensate particles, controlled by atmospheric circulation (e.g., Parmentier et al. 2013), may be inhomogeneous, e.g., asymmetric with respect to the substellar point. The patchy cloud scenario is motivated by a similar scenario proposed to explain a post-occultation phase shift for Kepler-7~b (Demory et al. 2013). A cloudy patch of the atmosphere should have different albedo than a clear patch of the atmosphere -- typically brighter -- therefore, patchy clouds can induce phase modulation at visible wavelengths. We do not separate planets having thick atmospheres and planets having thin atmospheres in the model, because we use a heat redistribution efficiency (see Section 3), rather than the surface pressure or the wind speed, as the primary model parameter.

\section{A General Model for Analyzing Visible-Wavelength Phase Curves}

We aim to construct a general model to describe the phase curve of an exoplanet observed at visible wavelengths. The phase curve measures $F_P/F_S$ as a function of the orbital phase, where $F_P$ is the flux of the planet and $F_S$ is the emission flux of the parent star. The planet's flux comes from disk-integrated reflection and thermal emission, as
\begin{equation}
F_P = F_R + F_T,
\end{equation}
where $F_R$ is the flux of stellar light reflected by the planet and $F_T$ is the thermal emission flux of the planet. The reflection component depends on the albedo of the planet, and the thermal emission component depends on the efficiency of heat redistribution, and a potential greenhouse effect if the planet has an atmosphere. 

The model is designed for planets that have circular orbits. The planetary systems for which the model would be applied to would have their primary and secondary transits well measured. The timing and duration of the transits constrain the orbital eccentricity (Kallrath \& Milone 1999; Barnes 2007; Ford et al. 2008; Kipping 2008; Demory et al. 2011). Therefore, one can confirm a planet to have a circular orbit before applying the model. Most {\it Kepler} planets that have measurement of the secondary eclipse depths have circular orbits (Esteves et al. 2014).

\subsection{Reflection Component}

We model the reflection component in the following two ways. The first possibility is that the reflection component is symmetric with respect to the occultation. This corresponds to the ``homogeneous surface'' and ``homogeneous atmosphere'' scenarios. The second possibility is that the reflection component is asymmetric with respect to the occultation. This corresponds to a planet with different reflectivities at different longitudes due to heterogeneous surface or atmosphere (i.e., the ``inhomogeneous surface'' and ``patchy cloud" scenarios). 

For symmetric reflection, we approximate the reflection component by the phase curve of a Lambertian sphere (i.e., all incident photons are isotropically scattered), scaled by a uniform geometric albedo ($A_g$) of the planet, namely
\begin{equation}
F_R^S(A_g) = F_S \bigg(\frac{R_P}{a}\bigg)^2 A_g \frac{1}{\pi} [\sin|\alpha| + (\pi-|\alpha|)\cos|\alpha|], \label{Lambert}
\end{equation}
where $\alpha$ is the phase angle of the planet that ranges from $-\pi$ to $\pi$ ($\alpha=0$ is at occultation and $\alpha=\pi$ is at transit), $R_P$ is the planet's radius, and $a$ is the semi-major axis. The Lambertian approximation is sufficient for current investigation because it is very difficult to extract a non-Lambertian component from a symmetric phase curve (Seager et al. 2000; Madhusudhan \& Burrows 2012). If the planet has a homogeneously reflecting atmosphere, the shape of its reflection phase curve will closely resemble the Lambertian phase curve (Seager et al. 2000; Cahoy et al. 2010). For an airless rocky planet, we have used the Hapke planetary regolith reflection model to compute the shape of its phase curve based on the method described in Hu et al. (2012), and confirmed that the resulting phase curve will be sufficiently approximated by a Lambertian one, as long as the particle size of the regolith is less than 100 $\mu$m.

For asymmetric reflection we consider some longitudes to have a high reflectivity and some longitudes to have a low reflectivity. One could envision a lava lake picture -- the molten surface has a lower reflectivity than the solid surface. One could also envision a ``hole-in-a-cloud'' picture -- the atmosphere is clear and poorly reflective at some longitudes near the substellar point and cloudy and highly reflective at other longitudes. We denote the low reflectivity as $r_0$ and the high reflectivity as $r_0+r_1$. Here a positive value for $r_1$ represents the increase in reflectivity caused by surface freezing or cloud formation. We further define $[\xi_1,\xi_2]$ as the local longitude\footnote{The local longitude is defined such that the substellar meridian is at the longitude of zero, the dawn terminator is at $-\pi/2$, and the dusk terminator is at $\pi/2$. The ``dawn'' and ``dusk'' are defined for a planet of prograde rotation.} range having low reflectivity, and then the local longitude ranges having high reflectivity are $[-\pi/2,\xi_1]$ and $[\xi_2,\pi/2]$. The reflection component is
\begin{equation}
F_R^A = F_R^S(2r_0/3) + F_S \bigg(\frac{R_P}{a}\bigg)^2 \frac{2r_1}{3} \frac{2}{\pi}\int_{\rm solid} \cos(\alpha-\phi)\cos(\phi)d\phi,
\label{Asymmetric}
\end{equation}
in which the first term is the Lambertian phase curve, characterized by $r_0$, and the second term contains integration over the high-reflectivity longitudes visible to observers, characterized by $r_1$. The integration is expressed in the observer's longitude ($\phi$) such that the observer is at the direction of $\phi=0$. The relationship between the observer's longitude and the local longitude is $\xi\equiv\phi-\alpha$. The exact expression for the second term as a function of $\xi_1$ and $\xi_2$ is explicitly calculated and given in Appendix \ref{Cloud}.

\subsection{Thermal Emission Component}

We calculate the thermal emission component in the {\it Kepler} band by multi-color blackbody emission based on a longitudinal distribution of temperature. The thermal emission photons can be either from the planet's surface, or from a certain pressure level in the planet's atmosphere, which we broadly refer to as the ``photosphere''. The phase dependency of the planet's thermal emission is computed by
\begin{equation}
F_T = R_P^2\int_{-\frac{\pi}{2}}^{\frac{\pi}{2}}\int_{-\frac{\pi}{2}}^{\frac{\pi}{2}}
B_K[T(\alpha, \theta, \phi)]\cos^2\theta\cos\phi\ d\theta d\phi\ , \label{therint}
\end{equation}
where $B_K$ is the Planck function integrated over {\it Kepler}'s bandpass, and the coordinates are specified by observer's latitude and longitude $(\theta,\phi)$, such that the observer is at the direction of $(\theta=0,\phi=0)$. The thermal emission component at each phase angle $\alpha$ is controlled by the corresponding temperature distribution $T(\theta, \phi)$.

The temperature distribution is determined by interaction between irradiation, heat redistribution, and radiative cooling. Atmospheric circulation models have been developed for irradiated gas giants (e.g., Showman \& Guillot 2002; Showman et al. 2009; Rauscher \& Menou 2010, 2012; Heng et al. 2011a, b; Perna et al. 2012; Showman et al. 2013; Dobbs-Dixon \& Agol 2013; Mayne et al. 2014; Rauscher \& Kempton 2014; Showman et al. 2014) and super Earths (e.g., Castan \& Menou 2011; Wordsworth et al. 2011; Heng \& Kopparla 2012; Kataria et al. 2014). Here we aim at fast calculation of the temperature distribution that would enable parameter exploration. Therefore, we adopt the semi-analytical model of Cowan \& Agol (2011), in which the atmosphere is mimicked by a rigidly rotating ``photosphere'' subject to irradiation and radiative cooling\footnote{As the thermal emission radiation in the {\it Kepler} band would almost entirely come from the equatorial region on the planet, a rotating photosphere is sufficient to describe the longitudinal variation of the photosphere temperature.}. 
%We also use this formulation of the rigidly rotating photosphere to simulate the temperature distribution on the planet's surface.
The temperature distribution is written as
\begin{equation}
T(\alpha, \theta, \phi) = f T_0 (\theta) {\cal P} (\epsilon, \xi) , \label{temp}
\end{equation}
in which $T_0$ is the temperature of the sub-stellar meridian, ${\cal P}$ is the thermal phase function that only depends on the local longitude and a heat redistribution efficiency ($\epsilon$), and $f$ is a scaling factor to account for a possible greenhouse effect. ${\cal P}$ is computed by solving Equation (10) in Cowan \& Agol (2011), 
\begin{equation}
\frac{d {\cal P}}{d\xi} = \frac{1}{\epsilon} (\max(\cos\xi,0)-{\cal P}^4),
\end{equation}
and the sub-stellar temperature is
\begin{equation}
T_0=T_S\bigg(\frac{R_S}{a}\bigg)^{1/2} (1-A_B)^{1/4} \cos(\theta)^{1/4}  ,\label{T0}
\end{equation}
where $T_S$ and $R_S$ are the stellar effective temperature and radius, respectively, and $A_B$ is the Bond albedo of the planet. 

In this toy model, the heat redistribution efficiency is defined as the ratio between the radiative timescale and the advective timescale, i.e.,
\begin{equation}
\epsilon = \tau_{\rm rad} \omega_{\rm adv}, \label{eps}
\end{equation}
where $\tau_{\rm rad}$ is the radiative timescale of the photosphere, and the advective frequency $\omega_{\rm adv}$ is $\omega_{\rm adv}\equiv\omega_{\rm photosphere}-\omega_{\rm orbit}$. $\omega_{\rm photosphere}$ and $\omega_{\rm orbit}$ are the angular velocities of the photosphere and the bulk part of the planet in the inertial frame of reference, respectively. When $|\epsilon|\gg 1$ the longitudinal variation of temperature will be small as heat redistribution is much more efficient than radiative cooling; when $|\epsilon|\ll 1$ the planet will be in local thermal equilibrium and the day-night contrast will be large. The sign of $\epsilon$, inherited from the sign of $\omega_{\rm adv}$, indicates the direction of the equatorial jets and the thermal phase shift: when $\epsilon>0$ the photosphere is super-rotating with respect to the planet's orbit, and the thermal phase shift is eastward (i.e., the peak of the thermal phase curve appears prior to the occultation, as in the case of hot Jupiter HD 189733b; e.g., Knutson et al. 2007); when $\epsilon<0$ the photosphere is sub-rotating with respect to the planet's orbit, and the thermal phase shift is westward (i.e., the peak of the thermal phase curve appears after the occultation). If the photosphere can be treated as a single atmospheric layer, the radiative timescale would be
\begin{equation}
\tau_{\rm rad} = \frac{c_p P}{g \sigma T_0^3}, \label{trad}
\end{equation}
where $c_p$ is the heat capacity of the atmosphere, $P$ is the pressure of the thermal emission photosphere, $g$ is the gravitational acceleration of the planet, and $\sigma$ is the Stephan-Boltzmann constant (Showman \& Guillot 2002).

Note that dynamical timescales other than the horizontal advection timescale may be important in controlling the temperature distribution on hot Jupiters; for example, in some cases, the vertical advection timescales and horizontal gravity wave propagation timescales exert a controlling influence on the day-night temperature pattern (Perez-Becker \& Showman 2013). In our above toy model, none of these dynamical processes are considered; $\omega_{\rm adv}$ should simply be viewed as a proxy for dynamical adjustment of the atmospheric temperature structure, by whatever mechanism.

The scaling factor $f$ in Equation (\ref{temp}) is included as a free parameter to mimic a possible greenhouse effect when there is an atmosphere. $f=1$ corresponds to the homogeneous surface and inhomogeneous surface scenarios without greenhouse effects, and $f>1$ corresponds to the planet that has an atmosphere with infrared-absorbing molecules. Such a scaling parameter is necessary because there could be infrared absorbers in the atmosphere, such as \ce{CO} and \ce{CO2}, which do not contribute significantly to the opacity at visible wavelengths. In other words, {\it Kepler} could potentially probe a deeper, and presumably warmer, layer on the planet compared to the layer that reflects stellar light. Using a uniform scaling factor to account for the possible greenhouse effect is of course coarse, as the radiative transfer processes in the atmosphere could be quite different from the dayside to the nightside (e.g., Burrows et al. 2008). However, without knowing the details of the atmospheric composition, a scaling factor would be best suited for the purpose of this model.

\subsection{Linking Reflection and Emission Components}

At this point we can link the longitudinal variation of reflectivity to the longitudinal variation of temperature. We assume that the pattern for lava or clouds follows a longitudinal distribution controlled by the local temperature of the surface or the atmosphere (see Appendix \ref{cloudmicrophysics} for justification). We introduce a single, physically motivated parameter $T_c$, called ``condensation temperature'' in the following, to describe the freezing temperature of the molten lava, or the condensation temperature of the condensable species in the atmosphere. When $T_0{\cal P} (\epsilon, \xi)>T_c$, the surface is molten, or the atmosphere is cloud free. When $T_0{\cal P} (\epsilon, \xi)<T_c$, the surface is solid, or the atmosphere is cloudy. $T_0{\cal P} (\epsilon, \xi)=T_c$ defines the longitudinal boundaries of the lava lake or the hole in the cloud, i.e., $\xi_1$ and $\xi_2$. We drop the $\cos(\theta)^{1/4}$ term in $T_0$ (Equation \ref{T0}) to determine $\xi_1$ and $\xi_2$, because any phase curve signal will be dominated by the atmospheric properties near the equatorial region that $\cos(\theta)\sim1$.
%Once we compute the temperature distribution with Equation (\ref{temp}), we determine $\xi_1$ and $\xi_2$ by comparing the temperature at each longitude to $T_c$.

Putting these pieces together, we find that for the symmetric reflection scenarios (i.e., lava planet and homogeneous atmosphere), the model is fully specified by three independent parameters: the Bond albedo of the planet ($A_B$), the heat redistribution efficiency ($\epsilon$), and the greenhouse factor $f$. The geometric albedo and the Bond albedo are linked by the phase integral; and since we work with the Lambert sphere assumption in this paper, $A_g=2A_B/3$. For the asymmetric reflection scenarios (i.e., homogeneous surface and patchy cloud), the model needs two additional parameters: the condensation temperature ($T_c$), and a reflectivity boosting factor ($\kappa$). The latter is defined as
\begin{equation}
r_1 \equiv \kappa r_0.
\end{equation}
$A_B$ and $\kappa$ are two independent parameters, from which $r_0$ and $r_1$ can be calculated as
\begin{eqnarray}
&& r_0= \frac{A_B}{1+\frac{2}{3}q'\kappa},\nonumber\\
&& r_1= \frac{\kappa A_B}{1+\frac{2}{3}q'\kappa} ,
\end{eqnarray}
where $q'$ is the phase integral given in Appendix \ref{Cloud} and Figure \ref{PhaseInt}. For physical plausibility, we verify each simulation to ensure $r_0+r_1\leq1$, and reject all attempted models that do not satisfy this criterion.

To summarize, our model computes the combined reflection and thermal emission from an irradiated exoplanet observed at visible wavelengths, based on three (assuming symmetric reflection) or five (assuming asymmetric reflection) parameters that describe physical processes on the planet. Our model covers a wide range of potential scenarios for exoplanets, with or without an atmosphere. 

\subsection{Fitting to Light Curves}

The semi-analytical model can be used to fit observed light curves. Typically, we do not include the primary transit, assuming that it has been used to derived key planetary and orbital parameters, but we do include the secondary occultation in the fit. We use the model of Mandel \& Agol (2002) for the shape of ingress and egress of the eclipse. This way, the phase curve characteristics including the occultation depth, the phase amplitude, and the phase offset can be directly derived from our fitting results.

Since our model computes phase curves very fast, we can use the Markov-Chain Monte Carlo (MCMC) method to explore the parameter space and determine the posterior parameter distribution of the three or five parameters from the phase curve observations. In practice, we calculate 2 Markov chains, each containing 1 million steps, for each planetary scenario, with the MCMC method implemented as in Haario et al. (2006). This number of steps is sufficient for the convergence of the Markov chains for all parameters in the cases studied here, and we validate the convergence by comparing the two chains using the standard Gelman-Rubin statistics ($R<1.01$ for all parameters; Gelman \& Rubin 1992). The first half of each chain is considered the ``burn-in'' period and removed from the final results. The physically allowed ranges and the prior distributions of the model parameters are: $A_B$ uniformly ranges in [0, 1]; $\epsilon$ uniformly ranges in $[-\infty, \infty]$; $f$ uniformly ranges in $[1, 2]$; $T_c$ uniformly ranges in [200, 3000] K; and $\kappa$ uniformly ranges in $[0, \infty]$. The actual ranges of parameters used in the fits to specific planetary scenarios may be narrower than these general ranges.

Our model does not require knowledge of the stellar radius. With our formulation, the planetary phase curve ($F_P/F_S$) depends on $R_P/R_S$ and $a/R_S$, both of which are uniquely derived from the primary transit (Seager \& Mallen-Ornellas 2003). In practice, the precision for the measurements of $R_P/R_S$ and $a/R_S$ is on the order of $1\%$ for Jupiter-sized planets whose transits and eclipses are detected by {\it Kepler} (e.g., Esteves et al. 2014), and the precision for the super-Earth-sized planet Kepler-10~b is better than $\sim3\%$ (Batalha et al. 2011). It is therefore legitimate to not propagate the uncertainties in $R_P/R_S$ and $a/R_S$ to the fitted parameters, when the uncertainties of the fitted parameters are much greater than $\sim1\%$. The model also depends on the stellar effective temperature. The stellar temperature is not cancelled out because of non-linear dependency of the Planck function on the temperature. For well-characterized {\it Kepler} stars, such as Kepler-7~b and Kepler-10~b, the precision on the stellar temperature is well within 1\%. We have tested our models and found 1\% change in the stellar temperature would produce negligible change in the phase curve.

In this paper we do not explicitly treat the doppler beaming or ellipsoidal effects to the phase curve, assuming that these effects can be effectively removed for transiting planets because relevant orbital parameters are known from the transits. The interaction between these effects and the atmospheric signatures will be discussed in a separate paper (Shporer \& Hu 2015).

\section{Applications to Kepler-7~b}

\begin{figure}
\begin{center}
 \includegraphics[width=0.45\textwidth]{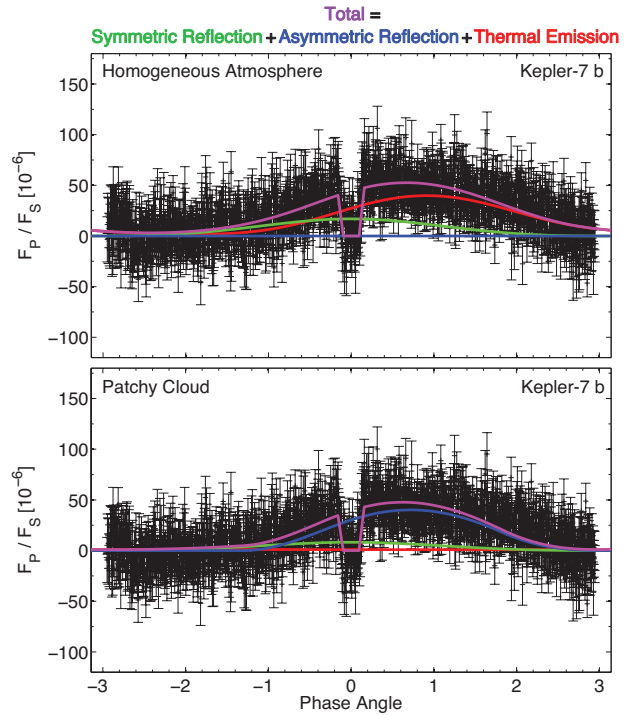}
 \caption{
The best-fit models for the phase curve of Kepler-7~b. The magenta lines are the modeled phase curves, and the other lines show contribution of thermal emission (red), symmetric reflection (green), and asymmetric reflection (blue). The upper-panel shows the best-fit model for a planet having a homogeneous atmosphere, and the modeled phase curve has significant contribution from a thermal emission component characterized by a hot spot shifted westward. By definition in this scenario the asymmetric reflection component is zero. The lower-panel shows the best-fit model for a planet having an atmosphere with patchy clouds, and the modeled phase curve has significant contribution from the asymmetric reflection component attributed to the clouds. The clouds concentrate on the west side of the substellar point, driven by a hot spot shifted eastward.
 }
 \label{K7Fit}
  \end{center}
\end{figure}

\begin{figure*}
\begin{center}
 \includegraphics[width=0.9\textwidth]{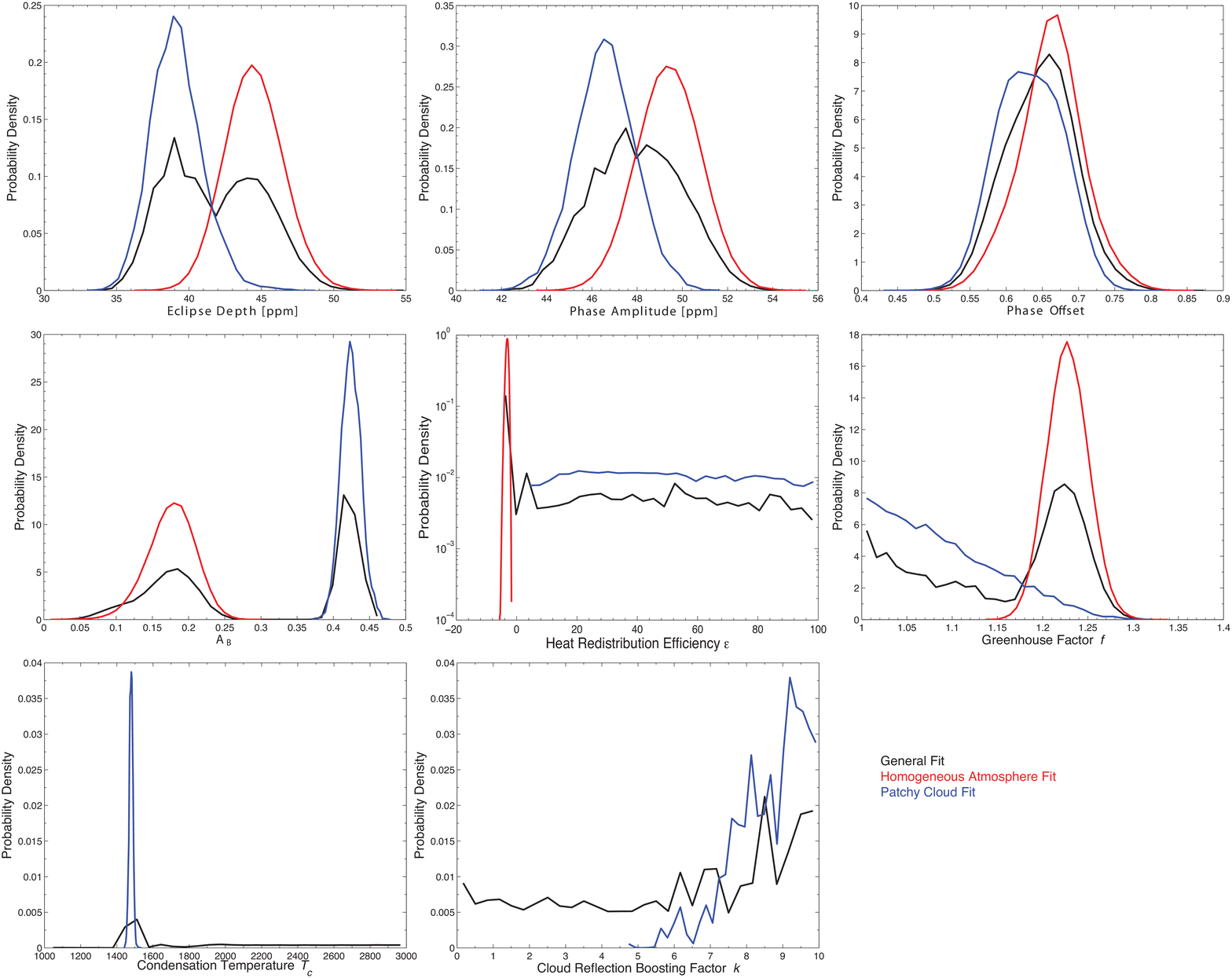}
 \caption{
Posterior probability distribution of the parameters for Kepler-7~b from MCMC simulations to fit the phase curve. Results from the general fit, the homogeneous atmosphere fit, and the patchy cloud fit are shown by different colors, and the color designation is tabulated on the lower right.
The posterior probability distribution resulted from the general fit shows bimodal solutions, and appears to be the combination of the results from the homogeneous atmosphere fit and the patchy cloud fit.
The homogeneous atmosphere fit tightly constrains the heat redistribution efficiency to $-3.2\pm0.5$, and the greenhouse factor to $1.23\pm0.02$.
The patchy cloud fit allows a positive $\epsilon$, but the phase curve yields no constraint on the exact value for $\epsilon$; instead, the phase curve well constrains the cloud condensation temperature.
% For the scenario with asymmetric reflection (patchy cloud), the posterior distribution of the derived lava/cloud boundary local longitudes ($\xi_1$ and $\xi_2$), and the posterior distribution of the derived reflectivity parameters ($r_0$ and $r_1$) are also shown. 
%The probability for a positive value of the heat redistribution efficiency (as predicted by eastward jets on a tidally synchronized planet) is 0.004 for the atmospheric scenario without patchy clouds. Both lava lake and patchy cloud scenarios allow a positive $\epsilon$, but the phase curve yields no constraint on the exact value for $\epsilon$; instead, the condensation temperature $T_c$ is fairly well constrained by the phase curve.
 }
 \label{K7PDF}
  \end{center}
\end{figure*}

\begin{figure*}
\begin{center}
 \includegraphics[width=0.9\textwidth]{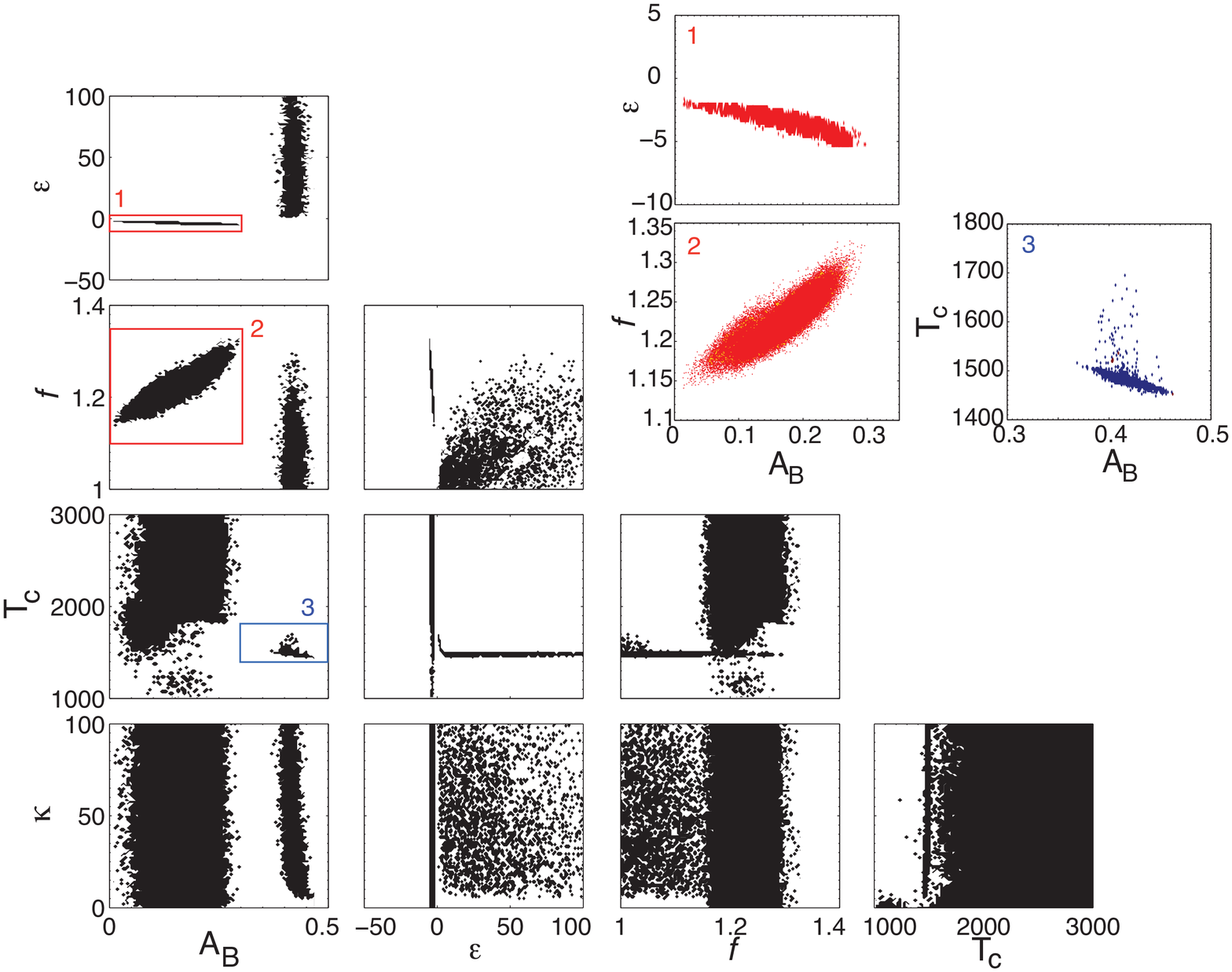}
 \caption{
Correlations between fitted parameters for Kepler-7~b. The results from the general fit are shown as black dots. 
The figures show two separate populations corresponding to the homogeneous atmosphere solution and the patchy cloud solution. 
Small Bond albedos ($A_B$), negative heat redistribution efficiencies ($\epsilon$), and greater-than-unity greenhouse factors ($f$) correspond to the homogeneous atmosphere solution; the other two parameters are unconstrained. The three constrained parameters are correlated, and their correlations are highlighted by inserted panel 1 and 2 that provide zoom-in views.
Large Bond albedos, positive heat redistribution efficiencies, well-constrained condensation temperatures ($T_c$) near 1500 K, and positive reflectivity boosting factors ($\kappa$) correspond to the patchy cloud solution. The Bond albedo and the condensation temperature are correlated and their correlations are highlighted by inserted panel 3.
%Three zoom-in views of the key correlations are shown as colored dots, with the red color corresponding to the correlations in the homogeneous atmosphere solution, and the blue color corresponding to the correlations in the patchy cloud solution. 
%The Bond albedo and the greenhouse factor are correlated for the homogeneous atmosphere fit, because in this case the phase curve is dominated by thermal emission and thus constrains the photosphere temperature.
 }
 \label{K7Cor}
  \end{center}
\end{figure*}

We now apply our model framework to analyze the visible-wavelength phase curves of exoplanets. {\it Kepler} has provided data to derive visible-wavelength phase curves for a number of Jupiter-sized exoplanets, and among these planets, Kepler-7~b has the best signal-to-noise ratios for the occultation depth and the phase offset (Demory et al. 2013; Esteves et al. 2014). Kepler-7~b has a circular orbit with an orbital eccentricity less than 0.02, the $3-\sigma$ upper limit derived from the light curve (Demory et al. 2011). As an example, we apply our model to analyze the phase curve of Kepler-7~b. We use the phase curve data of Demory et al. (2013) in this study. To focus on testing what we could learn from visible-wavelength phase curves, we only use the phase curve for the fit, and then consider the constraints of the occultation depths measured at longer wavelengths by Demory et al. (2013). We show our results in Table \ref{K7Para} and Figures \ref{K7Fit}, \ref{K7PDF}, and \ref{K7Cor}.

\begin{table*}[htdp]
\caption{Estimation of parameters for the gaseous planet Kepler-7~b, based on fitting to the observed phase curve. We use the photometry derived by Demory et al. (2013). In the general fit, all five parameters are allowed to vary in their physically plausible ranges. In the homogeneous atmosphere fit, the asymmetric reflection component is not included in the calculation, and then the model no longer depends on the cloud condensation temperature or the cloud reflection boosting factor. In the patchy cloud fit, we set $\epsilon>0$, assuming the planet to have super-rotating equatorial winds. Under such assumption, patchy clouds are required to explain the observed phase curve (see the text).
%The best-fitted model has $\chi^2=58.71$ and $\chi^2/dof=0.763$.
}
\begin{center}
\begin{tabular}{llllll}
\hline\hline
Parameter & General & Homogeneous Atmosphere & Patchy Cloud \\
\hline
\multicolumn{4}{l}{\it Fitted Parameters}\\
$A_B$ 		& $0.30_{-0.15}^{+0.13}$ & $0.18\pm0.03$ & $0.42\pm0.01$  \\
$\epsilon$ 	& $-3.3\sim62$ & $-3.2\pm0.5$ & $20\sim82$ \\
$f$ 			& $1.19_{-0.16}^{+0.05}$ & $1.23\pm0.02$ & $1.08^{+0.08}_{-0.05}$  \\
$T_c$ (K) 	         & $1476\sim2604$  & - & $1480\pm10$  \\
$\kappa$ 	         & $14\sim82$   & - & $13\sim69$  \\
\hline
\multicolumn{4}{l}{\it Derived Parameters}\\
Occultation Depth (ppm) & $41.5_{-3.2}^{+3.9}$ & $44.5\pm2.0$ & $39.0^{+1.8}_{-1.6}$ \\
Phase Amplitude (ppm)  & $48.0\pm2.1$ & $49.4\pm1.4$ & $46.6\pm1.3$ \\
Phase Offset (degree) & $37.2_{-3.0}^{+2.6}$ & $38.0_{-2.6}^{+2.4}$ & $36.2\pm2.7$ \\
$r_0$& $0.003\sim0.049$ & - & $0.014\sim0.072$ \\
$r_1$ & $0.15\sim0.94$ & - & $0.92\pm0.04$\\
$\xi_1$ (degree) & $-13.8\sim0$ & - & $-11.2\pm2.7$ \\
$\xi_2$ (degree) & $0\sim90$ & - & 90 \\
\hline
\multicolumn{4}{l}{\it Fit Quality}\\
Minimum $\chi^2/dof$  & 0.992 & 0.991 & 1.028 \\
BIC & 1340.8 & 1327.1 & 1387.1 \\
\hline\hline
\end{tabular}
\end{center}
\label{K7Para}
\end{table*}

A bimodal distribution for the fitted and derived parameters emerge when we apply the model framework to explain the phase curve of Kepler-7~b. The key feature of the observed phase curve is that the peak of planetary light occurs after the secondary occultation, i.e., a post-eclipse phase offset. Our model can generate this offset and provide a satisfactory fit to the observation (Figure \ref{K7Fit}). When all parameters are allowed to vary in their physically plausible ranges (i.e., the general fit), the posterior distribution shows two clearly separate parameter spaces that produce the fit to the observation (Figure \ref{K7PDF} and \ref{K7Cor}). The bimodal distribution is especially apparent for the Bond albedo, which can take values around either 0.2 or 0.4. Such a bimodal distribution indicates that two classes of models can be consistent with the observed phase curve.

The two classes of models correspond to the homogeneous atmosphere scenario and the patchy cloud scenario, respectively. To separate the two classes, we additionally perform two fits, each corresponding to one planetary scenario, by limiting the ranges in which the fitted parameters can vary. For the homogeneous atmosphere fit, we assume the asymmetric reflection component to be zero, which makes the cloud condensation temperature and the cloud reflection boosting factor dummy parameters. For the patchy cloud fit, we assume the planet to have super-rotating equatorial winds ($\epsilon>0$). Under such assumption, homogenous atmosphere can no longer provide a fit, and patchy cloud would have to be invoked to explain the post-eclipse phase offset. Table \ref{K7Para} shows the results of the two additional fits, and Figure \ref{K7PDF} compares the posterior distributions from the two separate fits with those from the general fit, and shows that the two separate fits correctly capture the two classes of solutions suggested by the general fit. 

While producing the same phase offset, the two classes of models result in slightly different occultation depths, because they differ in the shape of the phase curve. Table \ref{K7Para} shows that the derived occultation depths differ by 2-$\sigma$ between the homogenous atmosphere fit and the patchy cloud fit, and Figure \ref{K7PDF} shows that the general fit would give the combination of the above two fits, and therefore have much greater error bars. Comparing with Demory et al. (2013), the occultation depth from the patchy cloud fit is 1-$\sigma$ consistent, and that from the homogeneous atmosphere fit is only 2-$\sigma$ consistent. Similar bimodal distribution is also found for the posterior of the phase amplitude, but the two peaks differ to a lesser extent. This difference identified here highlights the importance to use appropriate models for the phase curve, even when the occultation depth bears most interest. A detailed look at the best-fit model phase curve (Figure \ref{K7Fit}) would reveal that the homogeneous atmosphere model has more thermal emission contribution than the patchy cloud model, and then the onset of the planetary light is smoother as the planet rotates from the nightside to the dayside. To fit the phase curve that contains the secondary occultation, the model automatically adjust the ``zero'' point of the planetary flux to seek the minimum $\chi^2$, which affects the determination of the occultation depth.

\subsection{Homogeneous Atmosphere}

The best-fit model for the homogeneous atmosphere scenario has thermal emission as the dominant source of planetary light. In this scenario, the post-occultation phase shift can be explained by a hot spot located on the west side of the substellar point (Figure \ref{K7Fit}). All three fitting parameters (the Bond albedo, the heat redistribution efficiency, and the greenhouse factor) are tightly constrained in this scenario (see the red lines in Figure \ref{K7PDF}). With only three parameters, the scenario achieves a superior goodness of fit indicated by $\chi^2$ and significantly better value for the Bayesian Information Criterion (BIC) than any other scenarios (Table \ref{K7Para}).

Assuming a homogeneous atmosphere, the visible-wavelength phase curve contains enough information to determine the three fitting parameters. The phase curve depends on these parameters non-linearly, and these parameters appear to be correlated but not fully degenerate (Figure \ref{K7Cor}). The most prominent correlation is between the greenhouse factor ($f$) and the Bond albedo ($A_B$), because the modeled temperature distribution is proportional to $f(1-A_B)^{1/4}$. These two parameters tend to be correlated, in order to maintain the temperature and then the thermal emission component (Figure \ref{K7Cor}). We also find that for a greater $A_B$, the model requires a greater (i.e., more negative) value for the redistribution efficiency $\epsilon$ (Figure \ref{K7Cor}). This is because when $A_B$ is greater, the symmetric reflection component becomes more significant and the thermal emission component becomes less significant. To keep the phase offset consistent with the observation, the offset of the thermal emission component alone would have to be greater, and therefore $|\epsilon|$ would have to be greater. Furthermore, when $|\epsilon|$ is greater, the temperature becomes lower at the dayside and higher at the nightside, which in turn would affect the occultation depth and the phase amplitude. In all, three ``model-independent'' observed quantities, the occultation depth, the phase amplitude, and the phase offset can uniquely determine the three fitting parameters. Once their values are found, we can separate the contribution from thermal emission and that from reflection, solely from the visible-wavelength phase curve.

Particularly for Kepler-7~b, we find that the heat redistribution efficiency ($\epsilon$) is smaller than zero by $6-\sigma$, implying that the advective frequency,  $\omega_{\rm adv} \equiv \omega_{\rm photosphere} - \omega_{\rm orbit}$ must have a large negative value, in order to explain the post-occultation phase shift. Therefore, the westward offset of the bright spot would seem to suggest that the air westward of the substellar point has the greatest temperature, and therefore that air at photospheric pressures travels westward in a synchronously rotating reference frame. This could occur on a synchronously rotating planet if the photospheric-level winds were westward at low latitudes.  However, to date, circulation models of highly irradiated, tidally locked exoplanets have generally predicted eastward equatorial winds and therefore eastward offsets of hot spots relative to the substellar point (e.g., Showman \& Guillot 2002, Cooper \& Showman 2005, Showman et al. 2008, 2009; Rauscher \& Menou 2010, 2012; Heng et al. 2011; Perna et al. 2012).  Such eastward jets were explained in a theory presented by Showman \& Polvani (2011), which shows that the day-night thermal forcing induces global-scale waves that transport prograde angular momentum to the equator, allowing such a so-called ``superrotating'' equatorial jet to emerge.  To date, no models of highly irradiated, synchronously locked planets have been published that exhibit a strong westward jet at the equator, as would seem to be needed to explain the westward offset in the Kepler-7~b phase curve in the homogeneous atmosphere scenario.

In principle, non-synchronous rotation could contribute to a hot spot offset. Circulation models of non-synchronously rotating hot Jupiters have been computed by Showman et al. (2009), Showman et al. (2014), and Rauscher \& Kempton (2014), considering rotation that is prograde (i.e., in the same direction as the orbital motion), with rotation periods both shorter and longer than the orbital period. Like their synchronously rotating counterparts, almost all of these models develop fast eastward equatorial jets -- fast enough to cause eastward motion in the synchronously rotating reference frame regardless of the rotation rate -- and therefore eastward hotspot offsets. That said, one slowly rotating non-synchronous simulation in Rauscher \& Kempton (2014) develops a ``westward'' equatorial jet that causes the thermal hotspot to be shifted west of the substellar point. In principle, planetary rotation that is retrograde (i.e., in the direction opposite the orbital motion) could also lead to thermal hotspot offsets of the correct sign to explain the Kepler-7~b data. However, given the short tidal spindown timescales for very hot Jupiters like Kepler-7~b, it is likely that the planet is close to synchronous rotation, in which case one would expect the existence of a superrotating jet and an eastward hotspot offset.

For Kepler-7~b, the dayside equilibrium temperature for the best-fit homogeneous atmosphere model is 1820 K, which is higher than the 3-$\sigma$ upper limit of the brightness temperature measured at the {\it Spitzer} 3.6 $\mu$m band (Demory et al. 2013). With the caveat that {\it Kepler} and {\it Spitzer} may probe different pressure levels and have different brightness temperatures, the homogeneous atmosphere scenario appears to be inconsistent with the {\it Spitzer} observations. Based on the atmospheric circulation model results and the {\it Spitzer} observations, the homogeneous atmosphere scenario is unlikely, which effectively makes the patchy cloud scenario to only plausible scenario for Kepler-7~b.

\subsection{Patchy Cloud}

The best-fit model for the patchy cloud scenario has the asymmetric reflection component as the dominant source of planetary light (Figure \ref{K7Fit}). In this scenario, reflective clouds located on the west side of the sub-stellar median can best explain the post-eclipse phase shift of Kepler-7 b. This is consistent with the explanation proposed by Demory et al. (2013) but our analysis offers more information on the atmosphere's properties. The Bond albedo is well constrained (Table \ref{K7Para}), and is much greater than that in the homogeneous atmosphere scenario. With a Bond albedo of $\sim0.4$, the dayside equilibrium temperature is consistent with the {\it Spitzer} observations.

The asymmetric reflection component is produced by a cloud distribution, in which the east side of the sub-stellar median is devoid of reflective clouds. To form such a cloud distribution, the model requires non-zero positive values for the heat redistribution efficiency ($\epsilon$) and the cloud reflectivity boosting factor ($\kappa$) (Table \ref{K7Para}). With a positive value for $\epsilon$, the east side is hotter than the west side; and if the cloud condensation temperature is suitable, condensation can only occur on the west side but not on the east side. This forms a ``hole'' in the cloud on the east side. Particularly for Kepler-7~b,  the cloud boundary would have to be located at $\sim10^{\circ}$ to the west of the substellar point to produce the observed phase offset (Table \ref{K7Para}). A positive value for $\epsilon$ indicates super-rotating jets that transport heat towards east, consistent with atmospheric circulation theories (Showman \& Guillot 2002; Showman et al. 2011). 

Also, to explain the phase offset, the cloudy part of the atmosphere must be more reflective than the cloud-free part, and our analysis shows that this reflectivity contrast would have to be quite significant to explain the phase curve of Kepler-7~b. We find that the cloud-free part of the atmosphere must be quite dark, having a reflectivity less than 7\%, and the cloudy part of the atmosphere must be quite bright, having a reflectivity greater than 90\%, at visible wavelengths (Table \ref{K7Para}).

The significant contrast in reflectivities is one of the main reasons why the model can constrain the Bond albedo. The Bond albedo cannot be too small, otherwise there would not be enough planetary light to explain the phase amplitude. Alternatively, the phase amplitude could be explained by additional thermal emission (i.e., by increasing the greenhouse factor $f$), but that would drive the overall phase shift towards the opposite direction than the asymmetric reflection. The Bond albedo cannot be too large also, because the hole in cloud has to be large enough (i.e., covering a significantly large part of the dayside), and dark enough for the significant contrast between the cloudy part and the cloud-free part. In all, the phase amplitude and the phase offset together put a tight constraint on the Bond albedo of the planet in the patchy cloud scenario.

In addition to the Bond albedo, fitting to the {\it Kepler} phase curve constrains the cloud condensation temperature and may imply the physical properties of the condensate species. The cloud condensation temperature is correlated with the Bond albedo, as $T_c\propto (1-A_B)^{1/4}$ (Figure \ref{K7Cor}). This is because the cloud condensation temperature needs to have a value between the maximum and the minimum temperature on the dayside, in order to produce a patchy cloud distribution. Caution should be exercised when comparing $T_c$ with the condensation curves of potential cloud-forming materials. We compare $T_c$ with the equilibrium temperatures in the model setup. However, the cloud may be located deep in the atmosphere at the pressure of 0.1-1 bars, and the true cloud condensation temperature may be higher than $T_c$. A more realistic range for the cloud condensation temperature would be between $T_c$ and $T_cf$, where $f$ is the derived greenhouse factor. For Kepler-7~b, we find this range corresponds to 1480-1730 K, using the values tabulated in Table \ref{K7Para}. The condensation curves of \ce{Fe}, \ce{Mg2SiO4}, \ce{MgSiO3}, and \ce{Cr} cross the inferred temperature range at the pressure of $10^{-3}-1$ bars for the solar abundances (Lodders \& Fegley 2006). \ce{Fe} and \ce{Cr} are strongly absorptive in the visible wavelengths, and thus cannot lead to the required high reflectivities. \ce{Mg2SiO4} and \ce{MgSiO3} are highly reflective (Sudarsky et al. 2003), and therefore are candidate cloud-forming materials for the atmosphere of Kepler-7~b.

Finally, the efficiency of heat redistribution cannot be sufficiently constrained by the phase curve. For Kepler-7~b, a non-zero positive value for $\epsilon$ is required, but the phase curve does not prefer a specific value for $\epsilon$ (Figure \ref{K7PDF}) - meaning that the phase curve is rather insensitive to the exact temperature distribution as long as a hot spot offset exists. This is because when $\epsilon$ is large, the longitudinal variation of temperature would be small (Cowan \& Agol 2011), and a slight adjustment of $T_c$ would be enough to keep the cloud boundary unchanged. This is different from the homogeneous atmosphere scenario, in which the phase offset directly depends on the hot spot offset and $\epsilon$ can be tightly constrained. In the patchy cloud scenario, due to the uncertainties of the cloud condensation temperature, the phase curve sets a $1-\sigma$ lower bound of 20 and does not yield an upper bound. In other words, the phase curve only requires a ``fairly significant'' heat redistribution, but cannot yield quatitative constraints on this parameter. 

\section{Applications to Kepler-10~b}

We derive a full phase curve of Kepler-10~b based on {\it Kepler} observations during the quarters 1 to 17. Kepler-10~b is a 1.4-R$_{\earth}$, 4.6-M$_{\earth}$ predominantly rocky planet (Batalha et al. 2011; Fogtmann-Schulz et al. 2014), making our result among the first revelation of any rocky exoplanets' phase curve signatures. This observation is made possible by continuous monitoring of the system by {\it Kepler} that brings down the error budget of photometry. We also benefit from the fact that the planet is hot enough to have significant thermal emission contribution to the {\it Kepler} band (Rouan et al. 2011), and the star is intrinsically quiet (Batalha et al. 2011). We apply our semi-analytical model to analyze the phase curve of Kepler-10~b.

\subsection{Data Reduction}

Our data reduction is similar to the one for Kepler-7~b presented in Demory et al. (2013). We use {\it Kepler} (Batalha et al. 2013) long-cadence simple aperture photometry (Jenkins et al. 2010) obtained during the quarters 1 to 17. We take into account the crowding matrix correction factor indicated in each FITS file on a quarter-per-quarter basis. We mitigate instrumental systematics by fitting the first four cotrending basis vectors (CBV) to each quarter using the PyKE software (Still \& Barclay 2012). We then normalize each quarter to the median. We account for photometric trends longer than four times the planetary orbital period by fitting a second-order polynomial to the out-of-eclipse data in the MCMC framework detailed below. We estimate and include the corrected noise the same way as in Demory et al. (2013). We find a nominal level (less than 10\%) of correlated noise throughout the dataset. 

\subsection{Model-Independent Analysis}

% The aim of our data analysis is to refine the transit parameters of Kepler-10~b and to characterize both the occultation and the phase-curve properties. 

Before isolating the planetary phase curve signal we search for all frequencies in the dataset to assess any risk of contamination. A typical Lomb-Scargle periodogram is not optimal in the case of datasets spanning long observations as slightly changing periodicities damp amplitudes in the power spectrum. To quantify how frequencies and amplitudes evolve in our dataset, we perform a wavelet transform analysis (e.g., Torrence \& Compo 1998) using the weighted wavelet Z-transform algorithm developed by Foster (1996). We do not detect any clear signature in the frequency/time spectrum, apart from the planet orbital signal. Kepler-10 is intrinsically quiet and any stellar activity remains nominal over the course of these {\it Kepler} observations. The frequency/time spectrum does not reveal quarter-dependent fluctuations.

We then conduct a model-independent Bayesian analysis of the entire dataset by employing the Markov Chain Monte Carlo (MCMC) implementation presented in Gillon et al. (2012). We assume a circular orbit, and set the occultation depth, phase-curve amplitude, phase-curve peak offset, period, transit duration, time of minimum light and impact parameter as jump parameters. We assume a simple trapezoid function for the occultation, and a Lambertian sphere modulation for the phase curve. We further assume a quadratic law for the limb-darkening (LD) and use $c_1=2u_1+u_2$ and $c_2=u_1-2u_2$ as jump parameters, where $u_1$ and $u_2$ are the quadratic coefficients. $u_1$ and $u_2$ are drawn from the theoretical tables of Claret \& Bloemen (2011) for the corresponding effective temperature and log $g$ values from Batalha et al. (2011). We add the two LD combinations $c_1$ and $c_2$ as Gaussian priors in the MCMC fit, based on the theoretical tables. We run two Markov chains of $10^5$ steps and assess their convergence using the statistical test from Gelman \& Rubin (1992). We also explore the effects of the data reduction parameters on deriving the transit parameters. We increase the CBV vectors to 8 and reduce and analyze the data separately, by pairs of quarters. During all these analyses variations, our MCMC fits result in individual transit parameter values within 1-$\sigma$ of the final Q1-Q17 values stated above.

We examine the robustness of the planetary phase-curve signal. We stack approximately three years of data, meaning that for stellar contamination to happen, stellar activity has to be phased exactly on the planetary orbital period of 0.8 days (or a multiple). Kepler-10 is an evolved star that is unlikely to have $v \sin i$ consistent with the short orbital period of Kepler-10~b. Kepler-10 b is not massive enough to cause any appreciable ellipsoidal or beaming components in the light curve. We therefore conclude that the orbital phase-curve is of planetary origin.

%\section{A Model-Independent View of the Light Curve}

\begin{figure}
\begin{center}
 \includegraphics[width=0.45\textwidth]{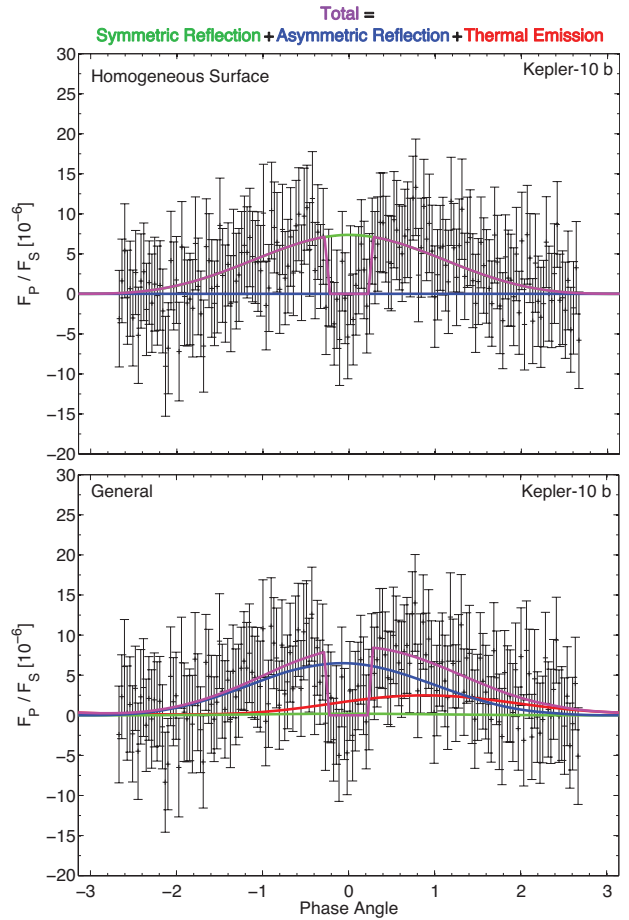}
 \caption{
Phase-folded light curve of Kepler-10~b system and examples of model fit. {\it Kepler} photometry obtained during the quarters 1 to 17 is used to derive this light curve, and key steps of data analysis are provided in \S~5. The light curve is binned to 200 bins for clarity. The magenta lines are the modeled phase curves, and the colored lines show contribution of thermal emission (red), symmetric reflection (green), and asymmetric reflection (blue). The upper panel shows the simplest model assuming a homogeneous surface. The model is dominated by the symmetric reflection component, and provides a good fit to the observed phase curve. The lower panel shows the best-fit model of the general fit, in which a thermal emission component from a hot spot shifted westward has significant contribution. Multiple models can provide adequate fit to the phase curve of Kepler-10~b.
 }
 \label{Fit}
  \end{center}
\end{figure}

We derive an occultation depth of $7.5\pm1.5$ ppm and a phase-curve amplitude of $8.5\pm1.2$ ppm (Figure \ref{Fit}). We do not detect a phase offset for Kepler-10~b, with the phase offset angle constrained to be $9\pm6$ degrees. Our value of the occultation depth is $1-\sigma$ consistent with previously reported values ($5.8\pm2.5$ ppm, Batalha et al. 2011; $9.9\pm1.0$ ppm, Fogtmann-Schulz et al. 2014)

The occultation depth translates to a brightness temperature at the {\it Kepler} band of $3220^{+90}_{-110}$ K if the planet's flux is from thermal emission, or a geometric albedo of $0.55\pm0.11$ if the planet's flux is from reflection. The phase curve magnitude, if solely attributed to reflection, corresponds to an effective geometric albedo of $0.63\pm0.09$, or a Bond albedo of $0.94\pm0.13$ for a Lambertian sphere. Given that the zero-albedo dayside-average equilibrium temperature of the planet is 2570 K, both thermal emission and reflection can have considerable contribution to the planet's emerging radiation.

The photometric measurements immediately outside the transits contain information about the planet's nightside. The transits occur from -20 to 20 degree (first contact) for this system, and we take a 20-degree interval on both sides of the transit to derive the planet's nightside emission flux. The result is a flux of $-0.7\pm1.2$ ppm, which places the $1-\sigma$ upper limit of the nightside temperature at 2270 K. Unlike Fogtmann-Schulz et al. (2014), our analysis does not yield a definitive nightside emission flux. Note that this flux constraint is sensitive to the limb-darkening parameterization of the star, as well as the interval size used in the estimate. In any case, the nightside brightness temperature must be much lower than the dayside brightness temperature.

%temperature cannot be directly constrained from the {\it Kepler} photometry. %From the phase curve magnitude alone, Kepler-10~b may or may not have efficient heat redistribution.

\subsection{Model-Assisted Analysis}

\begin{figure*}
\begin{center}
 \includegraphics[width=1.0\textwidth]{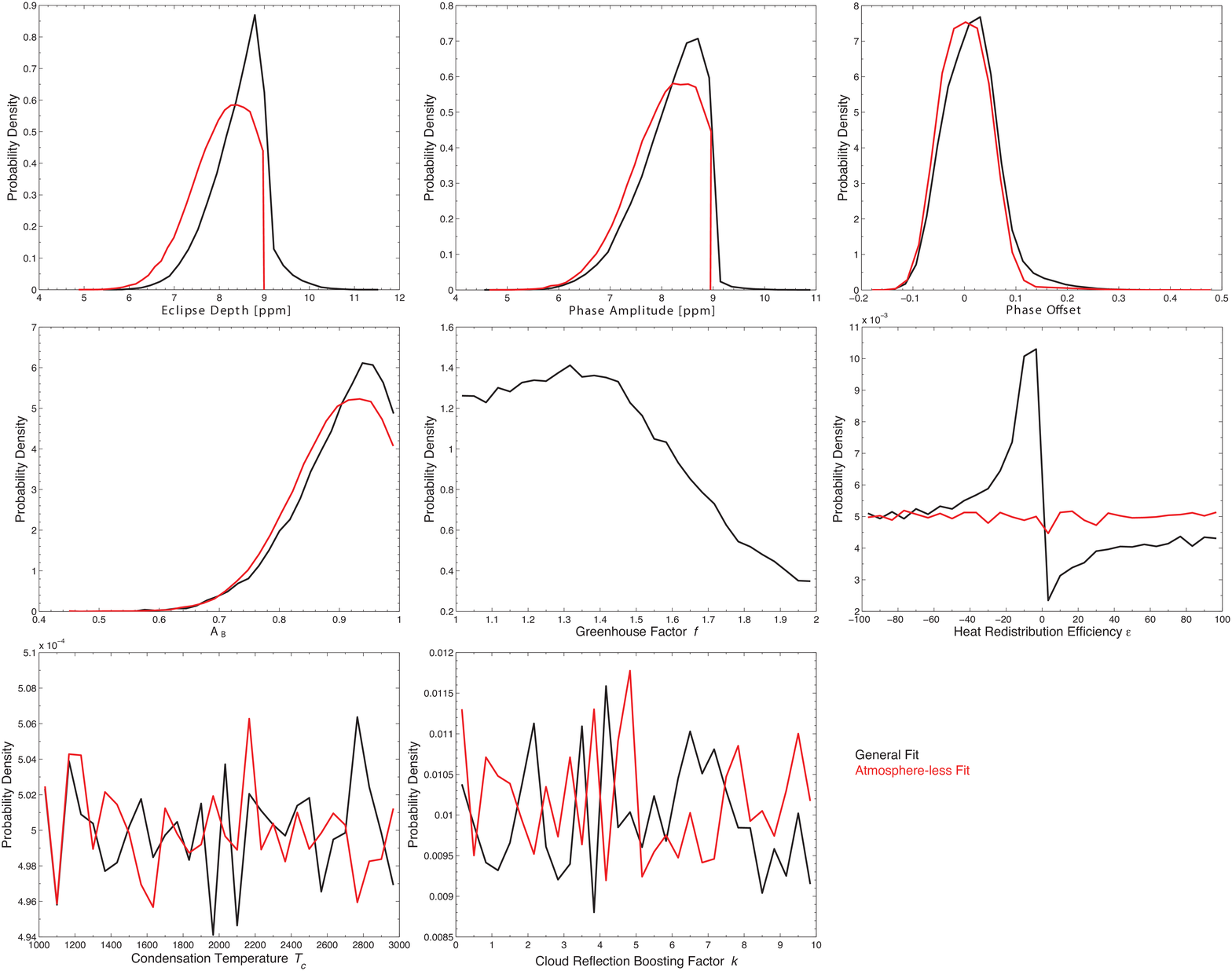}
 \caption{
 Posterior probability distribution of the parameters for Kepler-10~b from MCMC simulations to fit the phase curve. Black lines show the results of the general fit, and the red lines show the results of the atmosphere-less fit in which $f=1$ is assumed (i.e., no greenhouse effect).
The eclipse depth and the phase amplitude derived from the atmosphere-less fit have an upper limit of 9 ppm, and this upper limit corresponds to $A_B=1$. This upper limit does not exist when a greenhouse effect is allowed in the model. 
The general fit shows a slight preference to the model has shows a post-eclipse phase shift, driven by the negative heat redistribution efficiency. However, this phase offset is not statistically significant.
 }
 \label{PDF}
  \end{center}
\end{figure*}

%\subsection{Lava Planet}

We then apply our semi-analytical model to analyze the phase curve of Kepler-10~b. The eccentricity of the planet's orbit is consistent with zero (Batalha et al. 2011). We perform a general fit, in which all parameters are allowed to vary in their physically plausible range, and an atmosphere-less fit, in which the greenhouse factor is force to $f=1$, assuming that the planet does not an atmosphere. The atmosphere-less fit is motivated by the large bulk density of the planet, and the strong irradiation received by the planet (Batalha et al. 2011). The posterior distributions resulted from these fits are shown in Figure \ref{PDF}, and examples of model fits are shown in Figure \ref{Fit}.

We find that any model with a Bond albedo greater than 0.8 would provide an adequate fit to the observed phase curve, regardless of whether a greenhouse factor is used, or whether asymmetric reflection exists. Figure \ref{Fit} shows that the best-fit model phase curve is dominated by the reflection component. The only parameter constrained by the phase curve is the Bond albedo, and its value is $0.91^{+0.06}_{-0.10}$, consistent between the two fits. This value is also consistent with the model-independent estimate. This example shows that the constraints on the model parameters would be poor for rocky planets like Kepler-10~b, because the signal-to-noise ratio of the phase amplitude of the planet is only $7-\sigma$, much smaller than that of Kepler-7~b ($30-\sigma$).

Comparing the two fits, the phase curve does not yield additional constraints on whether the planet has an atmosphere. If Kepler-10~b does not have an atmosphere, its surface must be almost fully reflective (i.e., having a reflectivity of almost unity) in order to produce the magnitude of the secondary occultation and phase curve. Our estimate of the surface albedo is consistent with the high end of the range derived by Rouan et al. (2011). This is due to the new value of the occultation depth reported in this paper, which has a smaller error. The linear relationship between the Bond albedo and the occultation depth in Rouan et al. (2011) is still valid -- plugging-in our new value of the occultation depth to that equation would yield an estimate of the Bond albedo consistent with ours. Such a high reflectivity makes a solid surface possible. The dayside average equilibrium temperature would be 1720 K for a Bond albedo of 0.8, and only 1450 K for a Bond albedo of 0.9, and these temperatures are lower than the melting temperature of some materials such as aluminum oxide. Therefore it is unclear whether Kepler-10~b has a molten lava surface as suggested by Rouan et al. (2011); but if it does, the molten lava has to be highly reflective.

If Kepler-10~b has a thin atmosphere that provides moderate greenhouse effect ($f>1$), the thermal emission component can contribute to the planetary flux significantly. We find that the fit to the observed phase curve can be slightly improved, if a hot spot shifted to the west side of the substellar point is included (Figure \ref{Fit}). In the best-fit model, the shifted hot spot creates a post-eclipse phase offset that appears to be more consistent with the observations. The posterior distributions also show a preference to this class of solutions (Figure \ref{PDF}). However, due to the low signal-to-noise ratio of the phase offset, this improvement is marginal, and the constraint on the greenhouse factor is poor. 

\section{Discussion}

\subsection{Two Solutions for a Phase Offset}

When applying our semi-analytical model to study the phase curve of Kepler-7~b, we find two classes of solution can fit the phase curve equally well. The two classes of solution cannot be distinguish via usual statistical diagnostics like $\chi^2$ or the Bayesian Information Criterion. Here we generalize this dual solutions to any visible-wavelength phase curve that shows a significant phase offset for short-period exoplanets, and discuss how to break the degeneracy.

Take the post-eclipse phase offset for an example. Such an offset, qualitatively, means that the planet's brightness (from thermal emission, reflection, or both) of the west side of the substellar point is greater than that of the east side. If the phase curve is principally produced by reflection, the west side of the substellar point on the planet would have a greater albedo than the east side. If the phase curve is principally produced by thermal emission, the west side of the substellar point would be hotter than the east side. Either due to reflection or thermal emission, the phase shift indicates east-west transport on the planet. Here we see that, although the phase shift implies transport, there can be ambiguity in interpreting the direction of the phase shift, mainly due to entanglement of thermal emission and reflection.

Atmospheric circulation models may provide further constraints to break the degeneracy. For tidally-locked irradiated planets having a wide range of temperatures, atmospheric circulation models uniformly predict super-rotating equatorial winds and eastward shifted hot spot (e.g., Showman \& Guillot 2002; Showman et al. 2011). For these planets, a pre-eclipse phase offset would be driven by thermal emission, and the atmosphere can be homogeneously reflective; and a post-eclipse phase offset would be driven by asymmetric reflection due to formation of patchy clouds. 

Another way to break the degeneracy is to measure the planet's flux at a different wavelength, using different wavelength dependencies between the reflection and the thermal emission component. In the example of Kepler-10~b, a single measurement of the secondary occultation depth at mid-infrared wavelengths can rule out the scenarios in which thermal emission drives the phase offset. One caveat is that the opacities of the atmosphere can be very different between the visible and longer wavelengths, and the thermal emission at these wavelengths may probe the temperature at different pressure levels. Using planetary atmosphere chemistry and radiative transfer models can potentially mitigate this caveat (e.g., Hu \& Seager 2014). The joint observations at infrared wavelengths can only be done for few {\it Kepler} gaseous planets, due to the faintness of the sources. Looking ahead, TESS and PLATO will measure visible-wavelength phase curves of gaseous and rocky planets around nearby stars. For those systems, the {\it James Webb Space Telescope} can obtain their secondary occultation depths and provide great constraints to distinguish the two classes of models, and pinpoint the albedo and the thermal energy transfer of the planets.

Once this degeneracy can be broken with additional modeling or observational efforts, a lot more can be known about the planets. For example, if the phase offset is principally produced by thermal emission, the greenhouse factor $f$ could be tightly constrained by the phase curve. For a terrestrial exoplanet that may or may not have an atmosphere, a significant constraint of $f>1$ would indicate the planet must have an atmosphere. For another example, if the phase offset is principally produced by reflection, the cloud condensation temperature or the surface freezing temperature could be derived, which would indicate cloud or surface materials.

In this work we have assumed that all surface elements are Lambertian surfaces. For homogeneous scenarios, this assumption leads to $A_g=2A_B/3$. The real planetary atmospheres may be significantly backscattering. For example, if a planet has a semi-infinite atmosphere that is purely Rayleigh scattering, $A_g=3A_B/4$ (Dlugach \& Yanovitskij 1974). By neglecting backscattering we might overestimate the Bond albedo, by $10-15\%$. Correcting this bias leads to $1\%$ decrease in the estimate of the greenhouse factor $f$ for the homogeneous atmosphere solutions, or $1\%$ increase in the estimate of the cloud condensation temperature $T_c$ for the patchy cloud solutions.

In this paper we focus on planets that have circular orbits, which would include most of the planets whose phase curves can be measured by {\it Kepler} (Esteves et al. 2014). For completeness, we note that planets that have eccentric orbits may have phase offsets without heat redistribution (Cowan \& Agol 2011; Kataria et al. 2013). With heat redistribution, the phase curves of eccentric planets may show more complex patterns than the planets in circular orbits.

\subsection{Formation of Patchy Clouds}

The patchy cloud scenario is the leading explanation for the post-eclipse phase offset of Kepler-7~b. The inferred pattern of clouds follows a longitudinal distribution controlled by the local temperature of the atmosphere. Appendix \ref{cloudmicrophysics} shows that the condensation and sublimation are the dominant microphysical processes for 1-10 micron-sized particles at a temperature of 1000-2000 K and a pressure of 0.1-1 bars. Under these conditions, advection cannot transport cloud particles fast enough -- compared with evaporation -- to fill the ``hole'' in the cloud. Therefore, the patchy clouds may be condensational clouds that form instantaneously when the temperature falls below the condensation temperature.

In this picture, the thermal hot spot at 0.1-1 bars is the controlling dynamical pattern that affects the cloud appearance and the phase curve. However, at lower pressures, e.g. 1 mbar, the circulation would be dominated by symmetric flows from the substellar point to the anti-stellar point (e.g., Showman et al. 2008, 2009). As a result, the 0.1-1 bar thermal hot spot may correspond to a low temperature at 1 mbar, which could lead to another cold trap. The question is, can this low-pressure cold trap lead to cloud formation? We may consider this question by comparing the two timescales: the advective timescale of $10^4$ s (Appendix \ref{cloudmicrophysics}), and the vertical diffusive timescale. The latter can be estimated as $H^2/K_{\rm zz}$, where $H$ is the pressure scale height and $K_{\rm zz}$ is the vertical diffusivity. Using $H=50$ km and $K_{\rm zz}=5\times10^8$ cm$^2$ s$^{-1}$ (Parmentier et al. 2013), we estimate the vertical diffusive timescale to be $5\times10^4$ s. Therefore, on these planets, the condensable species can only be transported upward by a fraction of the pressure scale height for a typical advection timescale. The picture of the patchy clouds can then be, that on the nightside and the west side of the substellar point, the planet has a cold trap for the condensable species at the pressure of 0.1-1 bar, and due to the cold trap, the species is depleted at lower pressures. On the east side of the substellar point, the species can be transported upward, but this vertical transport would be effectively quenched by fast zonal winds and then clouds would not form at the low-pressure cold trap. Such a scenario can be studied by a model coupling atmospheric dynamics and cloud microphysics that includes condensation and sublimation.

The conditions to form the patchy clouds would thus be the availability of condensable material, a suitable atmospheric temperature, and a fast zonal wind. These conditions are met on gaseous planets having an equilibrium temperature of 1000-2000 K. As a result, patchy clouds featuring strong longitudinal variation of cloud coverage may be a common phenomena in the atmosphere of irradiated gaseous exoplanets. Although the cloud condensation temperature is fixed for a certain material, a wide range of stellar irradiation can still lead to patchy clouds, because the clouds can form at appropriate pressures corresponding to the condensation temperature. Our semi-analytical model has also been applied to two other {\it Kepler} planetary systems that show phase variation which cannot be due to orbital effects (i.e., doppler beaming and ellipsoidal effects). Similar post-eclipse phase offsets due to displaced patchy clouds are found, but diversities in the atmosphere and cloud reflectivities emerge (Shporer \& Hu 2015).

\section{Conclusion}

To analyze the visible-wavelength phase curves of exoplanets, we develop a semi-analytic model to calculate the reflection and thermal emission components of exoplanets as a function of orbital phase. The phase curve model depends on five parameters that characterize key energetic processes of the planets: the Bond albedo, the heat redistribution efficiency, the greenhouse effect factor (set to unity for atmosphere-less models), the condensation temperature (not used for symmetric reflection models), and the reflectivity boosting factor (not used for symmetric reflection models). The model tremendously simplifies the otherwise time-consuming computation for atmospheric circulation, cloud formation, and radiative transfer, which enables exploration of the full parameter space by the MCMC method. 

Our model framework applied to Kepler-7~b reveals a general degeneracy in interpreting the single-band phase curves of a hot exoplanet. If both reflection and thermal emission contributes to the planetary flux in the wavelength of interest, a potential phase shift can be attributed to a hot spot shifted to one direction in a clear atmosphere, or a hot spot shifted to the opposite direction in an atmosphere that has patchy clouds or on a surface that has inhomogeneous reflectivities. For Kepler-7~b we find both scenarios fit the visible-wavelength phase curve fairly well, despite vastly different atmospheric circulation regimes they imply. For Kepler-7~b, atmospheric circulation models and previous {\it Spitzer} observations rule out the model attributing the phase offset to the thermal emission component, and make the patchy cloud scenario the only plausible explanation. Our model can therefore derive strong constraints on the reflectivities of the clear part and the cloudy part of the atmosphere, and the location of the cloud boundary. We also present a full phase curve of Kepler-10~b, and improved measurements of the occultation depth and the phase magnitude. We do not detect a significant phase offset in the phase curve of Kepler-10~b. Due to the low signal-to-noise ratio of its phase amplitude, our model can only constrain the Bond albedo of the planet. In all, the phase curve analysis is a powerful tool to investigate the exoplanets discovered by {\it Kepler} and those to be discovered by future transit survey missions. This ``cost-free'' information from current planet searching strategy will yield highly valuable constraints on the planets' characteristics.

\acknowledgments

RH thanks Avi Shporer for providing the program to compute ingress and egress shapes, Tiffany Kataria for making available preliminary 3D atmospheric circulation simulations, Edwin Kite for discussion on the molten lava planet, Feng Tian for discussion on the atmospheric escape of rocky planets, and Wesley Traub and Yuk L. Yung for helpful discussions. This paper includes data collected by the Kepler mission. Funding for the Kepler mission is provided by the NASA Science Mission directorate. Support for RH's work was provided in part by NASA through Hubble Fellowship grant \#51332 awarded by the Space Telescope Science Institute, which is operated by the Association of Universities for Research in Astronomy, Inc., for NASA, under contract NAS 5-26555. Part of the research was carried out at the Jet Propulsion Laboratory, California Institute of Technology, under a contract with the National Aeronautics and Space Administration. NKL's work was performed in part under contract with the California Institute of Technology (Caltech) funded by NASA through the Sagan Fellowship Program executed by the NASA Exoplanet Science Institute.

\appendix

\section{Phase Function Associated with Asymmetric Distribution of Condensed Particles}

\label{Cloud}

The local longitude range on the planet illuminated by the star is $\big[-\frac{\pi}{2},\frac{\pi}{2}\big]$. For the scenarios with asymmetric reflection, the local longitudes ranging in $[\xi_1,\xi_2]$ have a lower reflectivity than other longitudes on the planet. $\xi_1$ and $\xi_2$ are set by comparison between the local temperature and the freezing temperature of the surface or the condensation temperature of the condensable species in the atmosphere, and they must also obey $-\frac{\pi}{2}\leq\xi_1\leq\xi_2\leq\frac{\pi}{2}$. 

The phase function of the reflected light associated with the additional reflection, which is in addition to the reflected light by a homogeneous surface or atmosphere, is
\begin{equation}
\Phi = \frac{2}{\pi}\int_{\rm solid}\cos(\alpha-\phi)\cos\phi d\phi , \label{eqp1}
\end{equation}
in which the integration covers the longitudes that have the high reflectivity and that can be seen along the line of sight. Given the relationship between the local longitude and the observer longitude, $\xi\equiv\phi-\alpha$, the phase function can also be integrated as
\begin{equation}
\Phi = \frac{2}{\pi}\int_{\rm solid}\cos\xi\cos(\xi+\alpha) d\xi . \label{eqp2}
\end{equation}
Define $\alpha'=-\alpha$, and Equation (\ref{eqp2}) with respect to $\alpha'$ takes the same form as Equation (\ref{eqp1}) with respect to $\alpha$. It is convenient to evaluate the phase function in $\alpha'$ and the exact formulae for this integral for the phase function are given as follows.

For $-\pi\leq\alpha'\leq0$,

If $-\frac{\pi}{2}\leq\alpha'+\frac{\pi}{2}\leq\xi_1$,
\begin{equation}
\Phi = \frac{1}{\pi} [\cos\alpha'(\pi+\alpha') + \sin(\pi+\alpha')] ;
\end{equation}
If $\xi_1\leq\alpha'+\frac{\pi}{2}\leq\xi_2$,
\begin{equation}
\Phi = \frac{1}{\pi} \{\cos\alpha'(\frac{\pi}{2}+\xi_1) + \frac{1}{2}[\sin(\pi+\alpha')+\sin(2\xi_1-\alpha')]\} ;
\end{equation}
If $\xi_2\leq\alpha'+\frac{\pi}{2}\leq\frac{\pi}{2}$
\begin{equation}
\Phi = \frac{1}{\pi} [\cos\alpha'(\pi+\alpha'+\xi_1-\xi_2) + \sin(\pi+\alpha') + \cos(\xi_1+\xi_2-\alpha')\sin(\xi_1-\xi_2)] ;
\end{equation}

For $\pi\geq\alpha'\geq0$,

If $-\frac{\pi}{2}\leq\alpha'-\frac{\pi}{2}\leq\xi_1$,
\begin{equation}
\Phi = \frac{1}{\pi} [\cos\alpha'(\pi-\alpha'+\xi_1-\xi_2) + \sin\alpha' + \cos(\xi_1+\xi_2-\alpha')\sin(\xi_1-\xi_2)] ;
\end{equation}
If $\xi_1\leq\alpha'-\frac{\pi}{2}\leq\xi_2$,
\begin{equation}
\Phi = \frac{1}{\pi} \{\cos\alpha'(\frac{\pi}{2}-\xi_2) + \frac{1}{2}[\sin\alpha'-\sin(2\xi_2-\alpha')]\} ;
\end{equation}
If $\xi_2\leq\alpha'-\frac{\pi}{2}\leq\frac{\pi}{2}$,
\begin{equation}
\Phi = \frac{1}{\pi} [\cos\alpha'(\pi-\alpha') + \sin\alpha'] .
\end{equation}

\begin{figure}[htdp]
\begin{center}
 \includegraphics[width=0.45\textwidth]{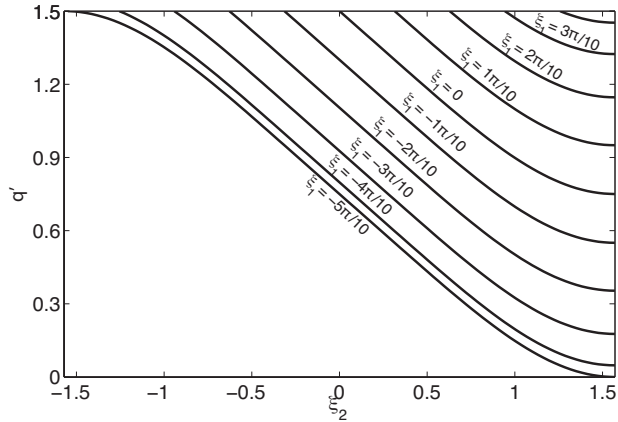}
 \caption{
The phase integral associated with the frozen surface or the condensed particles ($q'$) as a function of $\xi_1$ and $\xi_2$.
 }
 \label{PhaseInt}
  \end{center}
\end{figure}

We also need to calculate the contribution of the additional asymmetric reflection to the planet's Bond albedo. This is done by further integrating the phase function with respect to $\alpha$. The phase integral is
\begin{equation}
q'(\xi_1,\xi_2) = \int_{-\pi}^{\pi} \Phi \sin|\alpha| d\alpha .
\end{equation}
The value of $q'$ as a function of $\xi_1$ and $\xi_2$ is evaluated numerically and shown in Figure \ref{PhaseInt}. For example, if cloud particles cover the whole substellar hemisphere ($\xi_2=\xi_1$), $q'=3/2$, which gives the classical result of a Lambertian sphere. The Bond albedo of the planet is
\begin{equation}
A_B = r_0 + \frac{2}{3}q'(\xi_1, \xi_2)r_1 . \label{AB}
\end{equation}

\section{Microphysical consideration of the patchy clouds}

\label{cloudmicrophysics}

To simplify the lava lake scenario and the patchy cloud scenario, we assume that the pattern for lava or clouds follows a longitudinal distribution controlled by the local temperature of the surface or the atmosphere. For the patchy clouds, this is equivalent to saying that condensation and sublimation are the dominant microphysical processes, while precipitation and advection are relatively unimportant. Here we show that this assumption is likely to be valid for 1-10 micron-sized particles at a pressure level of 0.1 bars with a temperature of $\sim1000$ K. With this assumption, a non-uniform longitudinal distribution of temperature, as a result of energy redistribution, can lead to a non-uniform longitudinal distribution of cloud particles. 

For the patchy cloud scenario we have assumed that condensation and sublimation is controlled solely by the local temperature. To investigate when this ``local cloud'' assumption is valid, we compare the following timescales:
\begin{itemize}
\item Advective timescale $\tau_{\rm adv}\sim R_P/u$, where $R_P$ is the radius of the planet and $u$ is the equatorial wind speed. $u$ would be on the order of 1 km s$^{-1}$ for irradiated exoplanet atmospheres (e.g., Showman \& Guillot 2002). For Kepler-10~b $\tau_{\rm adv}\sim10^4$ s, and for Kepler-7~b it would be longer.
\item Condensation/sublimation timescale $\tau_{\rm cond}\sim\frac{D_p^2\rho_p k_bT}{12Km(p-p_{\rm sat})}$, where $D_p$ is the particle diameter, $\rho_p$ is the density of the condensed material, $k_b$ is the Boltzmann constant, $T$ is the dayside temperature, $K$ is the diffusivity of the atmosphere, $m$ is the molecular mass of the condensable species, and $p-p_{\rm sat}$ is the difference between the partial pressure and the saturation vapor pressure of the condensable material. The estimation formula is derived from the condensational growth equation of atmospheric aerosols for perfect accommodation and the continuum regime\footnote{Condensational growth will be in the continuum regime when particle radii are considerably greater than the mean free path of the condensable gas. We have estimated the mean free path to be 0.1 $\mu$m for a gas with a molecular mass of 100 at 2000 K. Therefore the particles of our concern, whose radii are in the order of 1 $\mu$m or greater, would satisfy the continuum regime. For the same reason, gravitational settling would also be in the continuum regime.} (Seinfeld \& Pandis 2006). Note that the condensation timescale is a strong function of the particle size and the partial pressure of the condensable species. Estimates of the condensation timescale using typical values are shown in Table 2.
\begin{table}[htdp]
\label{taucond1}
\caption{Condensation timescale of a potentially condensable species in the atmosphere. The typical values are taken as $\rho_p=3\times10^3$ kg m$^{-3}$ (enstatite), $T=2000$ K, $D=6\times10^{-5}$ m$^2$ s$^{-1}$, $m=1.7\times10^{-25}$ kg, and atmosphere pressure $P=10^4$ Pa.}
\begin{center}
\begin{tabular}{l|cc}
										& $D_p=10^{-6}$ m & $D_p=10^{-5}$ m \\
										\hline
$p-p_{\rm sat}=10^{-6}$ P		& 70 s					& $7\times10^3$ s \\
$p-p_{\rm sat}=10^{-8}$ P		& $7\times10^3$ s	& $7\times10^5$ s \\
\end{tabular}
\end{center}
\label{default}
\end{table}
\item Gravitationally settling timescale $\tau_{\rm settle}\sim \frac{H}{v_{\rm settle}}$, where $H$ is the atmospheric scale height and $v_{\rm settle}$ is the settling velocity. At $10^4$ Pa, the settling velocity for a $1~\mu m$ sized particle is in the order of 0.01 m s$^{-1}$, and the atmospheric scale height on Kepler-10~b is about 30-80 km. We estimate $\tau_{\rm settle}\sim3-8\times10^8$ s. This value is greater by a factor of a few for Kepler-7~b. For $10~\mu m$ sized particle $\tau_{\rm settle}$ would be smaller by 2 orders of magnitude.
\end{itemize}

Based on these timescale estimates, we find it possible to have $\tau_{\rm cond}< \tau_{\rm adv} < \tau_{\rm settle}$ for particles of 1-10 micron radius at the pressure level of $\sim0.1$ bars in the atmospheres of Kepler-10~b and Kepler-7~b. That is to say, formation and sublimation of a potential condensable species is in local equilibrium, i.e., the amount of condensate particles only depends on the local temperature. In comparison, horizontal and vertical transport can be much less efficient. These estimates also suggest that the particle size could be greatly limited by advection in the atmospheres having strong zonal winds, instead of limited by vertical transport as for some brown dwarfs (e.g., Cooper et al. 2003).
%$\tau_{\rm cond}\ll \tau_{\rm adv}$ indicates that it is physically plausible to have a non-uniform longitudinal distribution of cloud particles in the atmosphere of Kepler-10~b, controlled by the temperature distribution.

\end{document}